\documentclass[1p,number,12pt]{elsarticle}
\usepackage[utf8]{inputenc}
\usepackage[T1]{fontenc}
\usepackage{lmodern}
\usepackage[english]{babel}
\usepackage{amsmath}
\usepackage{amssymb}
\usepackage{graphicx}
\usepackage{wrapfig}
\usepackage{float}
\usepackage{xcolor}
\usepackage{algorithm}
\usepackage{algorithmic}
\usepackage{multirow}
\usepackage{bbm}

\usepackage{fancyhdr}
\newcommand{\specialcell}[2][c]{%
  \begin{tabular}[#1]{@{}c@{}}#2\end{tabular}}
  
\begin{document}

\begin{frontmatter}
\title{A New Estimator of Intrinsic Dimension Based on the Multipoint Morisita Index}
\author{Jean GOLAY and Mikhail KANEVSKI}
\address{Institute of Earth Surface Dynamics, Faculty of Geosciences and Environment, University of Lausanne, 1015 Lausanne, Switzerland. Email: jean.golay@unil.ch. Phone: +41 21 692 35 41}

\begin{abstract}
The size of datasets has been increasing rapidly both in terms of number of variables and number of events. As a result, the empty space phenomenon and the curse of dimensionality complicate the extraction of useful information. But, in general, data lie on non-linear manifolds of much lower dimension than that of the spaces in which they are embedded. In many pattern recognition tasks, learning these manifolds is a key issue and it requires the knowledge of their true \textit{intrinsic dimension}. This paper introduces a new estimator of intrinsic dimension based on the multipoint Morisita index. It is applied to both synthetic and real datasets of varying complexities and comparisons with other existing estimators are carried out. The proposed estimator turns out to be fairly robust to sample size and noise, unaffected by edge effects, able to handle large datasets and computationally efficient.  
\end{abstract}

\end{frontmatter}

\section{Introduction}
The $21^{st}$ century is more and more data-dependent and, in general, when collecting data for a particular purpose, it is not known which variables matter the most. This lack of knowledge leads to the emergence of high-dimensional datasets characterized by redundant features which artificially increase the volume of data to be processed. As a result, the empty space phenomenon \cite{Scott83} and the curse of dimensionality \cite{Bell61} make it challenging to conduct pattern recognition tasks such as clustering and classification.\\ 

The goal of Dimensionality Reduction (DR) \cite{LeeVer07,Burges09}, sometimes called manifold learning, is to address this issue by mapping the $N$ sampled data points into the lower dimensional space where they truly lie. Such a space is often considered as a manifold of intrinsic dimension $M$\footnote{In Physics, mainly, $M$ is often referred to as the degrees of freedom of data} embedded in a Euclidean space of dimension $E$ with $E \geq M$. $E$ equals the number of variables of a dataset and the Intrinsic Dimension (ID) of a manifold is equal to the theoretical ID of the data. If a manifold is space-filling, its dimension $M\approx E$. In contrast, if the Euclidean space is partially empty, $M <E$. The optimality of DR greatly depends on the accuracy of ID estimates. An underestimation of the theoretical ID will result in the implosion of the data manifold and information will be irreparably lost. On the contrary, an overestimation will lead to noise in the final mapping. From an application perspective, DR can be used to produce low dimensional syntheses of high dimensional datasets \cite{Ten00} and as a preprocessing tool for supervised learning \cite{Len00,KanPoz09} and data visualization \cite{Kaski11}.\\  

DR methods perform variable transformations to capture the complex dependencies which generate redundancy within datasets. Nevertheless, it is often important not to recast data. The Fractal Dimension Reduction (FDR) algorithm \cite{Trai00,Zha02,Sousa07,Trai10} was designed to this end. The fundamental idea is to remove from a dataset all the variables which do not contribute to increasing its ID. FDR can also be adapted to supervised feature selection methods \cite{Mo12}. The goal is then to reject irrelevant or redundant variables (or features) according to a prediction task (i.e. regression or classification). Although ID estimation lies at the core of FDR, more traditional unsupervised \cite{Dy04,Dy08,Ale13} and supervised \cite{Aha96,Jain97,Blum97,Mol02,Guy03,Guy06} feature selection methods do not consider it. It has, however, a great potential in speeding up search strategies, such as those used in \cite{Mar63,Pud94,Vieira07,Wang07}.\\

These different approaches highlight that ID estimation is a fundamental problem when dealing with high-dimensional datasets. Unfortunately, ID estimators \cite{Cama03,Lev04} suffer from the curse of dimensionality as well. Their overall performance depends on many factors (to various degree), such as the number of data points, the theoretical ID of data and the shape of manifolds. The present research deals with a new ID estimator in order to provide a solution to the problems raised by these factors. It is based on the recently introduced multipoint Morisita index ($m$-Morisita)\cite{Go14,Hul90,Mori59}. The $m$-Morisita index is a measure of global clustering closely related to the concept of multifractality and, so far, it has been successfully applied within the framework of (2-dimensional) spatial data analysis \cite{Go14,Hul90}.\\

The paper is organized as follows: In Section \ref{IDest}, traditional fractal-based and maximum likelihood methods of ID estimation are presented. Section \ref{Mindex} derives a new ID estimator from the $m$-Morisita index and introduces a new algorithm for its application to high-dimensional datasets. Section \ref{appcomp} is devoted to comparisons between the proposed estimator and those of Section \ref{Mindex}. Their behaviour regarding sample sizes, noise and the dimension of manifolds is analysed. A special attention is also paid to their bias and variance by using Monte-Carlo simulations and real world case studies from the UCI machine learning repository are examined. Finally, conclusions are drawn in Section \ref{clus}.

\section{Existing Methods}\label{IDest}
Many ID estimation methods have been proposed \cite{Cama03,Lev04,Jain88,The90,Keg02} and they can be roughly divided into projection (e.g. PCA) and geometric methods (e.g. fractal, nearest-neighbor and maximum likelihood methods). This section focuses on fractal-based and maximum likelihood estimators. They are commonly used in a wide range of applications and they generally provide non-integer values as ID estimates. 

\subsection{Fractal-Based Estimation Methods}
The word \textit{fractal} was first coined by B. Mandelbrot \cite{Mand83} to describe scale-invariant sets. At small scales $\delta$, for a given point pattern, one has that:\begin{equation} n_{box}(\delta) \propto \delta^{-D_0}\end{equation}where $n_{box}(\delta)$ is the number of grid cells necessary to cover the whole pattern and $D_0$ is known as the box-counting dimension \cite{Mand83,lov86,Smith89}. In practical applications, due to its simplicity, $D_0$ often replaces the Hausdorff dimension $D$ (or fractal dimension) and it can be proved that $D_0$ is an upper bound of $D$ \cite{Fal03}.\\ 

In complex cases, the scaling behaviour of the moments of point distributions cannot be fully characterized by only one fractal dimension and a full spectrum of generalized dimensions, $D_q$, is required. Such distributions are referred to as being multifractal \cite{Hent83,Grass832,Pal87,tel89}. $D_q$ is generally obtained by using a generalization of the box-counting method \cite{Hent83,Grass832,Pal87,Vic93} based on Rényi's information, $RI_{q}(\delta)$, of $q^{th}$ order \cite{Renyi70}. The central scaling law of this approach can be written as follows for $q\neq 1$:\begin{equation}\label{renyione} 
\exp(RI_{q}(\delta)) \propto \delta^{-D_{q}}
\end{equation}where \begin{equation}\label{Rinfo} 
 RI_{q}(\delta) = \frac{1}{1-q} \ \log(\sum_{i=1}^{n_{box}(\delta)} p_{i}(\delta)^q)
\end{equation}In this last equation, $p_{i}(\delta) = n_i/N$ is the value of the probability mass function in the $i^{th}$ grid cell of size $\delta$ ($n_i$ is the number of points falling into the $i^{th}$ cell) and $q \in \mathbb{R}\backslash\lbrace-1\rbrace$. Finally, one has that:\begin{equation} \label{Dq_cla} 
D_{q} = \lim_{\delta \to 0}\frac{RI_{q}(\delta)}{\log(\frac{1}{\delta})}
\end{equation}and\begin{align}
\lim_{q \to 1}D_q &= df_{i}\\
 D_2 &= df_{cor}
\end{align}
where $df_{i}$ and $df_{cor}$ are, respectively, the information dimension \cite{Renyi56,Hent83} and the correlation dimension \cite{Grass831}.\\ 

Usually, $df_{cor}$ is computed with the Grassberger-Procaccia (GP) algorithm \cite{Grass831}. This algorithm is designed to better take advantage of the range of available pairwise distances between points. It can be introduced as follows: at small scales, for a point set, $X_N=\lbrace x_1,\ldots ,x_N\rbrace$, one has that\begin{equation}
C(\delta) \propto \delta^{df_{cor}}
\end{equation}where\begin{equation} 
C(\delta) = \frac{2}{N(N-1)} \ \sum_{i=1}^N\sum_{j=i+1}^N \mathbbm{1}_{\left\lbrace  \Vert x_i - x_j\Vert \leq \delta \right\rbrace}   \end{equation}with $\mathbbm{1}$ being an indicator function and $df_{cor}$ can be expressed as\begin{equation}
df_{cor} = D_2 = \lim_{\delta \to 0} \frac{\log(C(\delta))}{\log(\delta)}
\end{equation}

The available values of $RI_q(\delta)$ and $\log(C(\delta))$ depend on the data resolution. A commonly used method for estimating $D_q$ consists in plotting $RI_q(\delta) \ vs \ \log(\delta^{-1})$ for a chosen scale interval. The final estimate is then the slope of the linear regression fitting the linear part of the resulting chart. The procedure is the same for the GP algorithm, except that $df_{cor}$ and $\log(C(\delta))$ replace, respectively, $D_q$ and $RI_q(\delta)$. Eventually, both $D_q$ (in general $0\leqslant q\leqslant 2$) and $df_{cor}$ can be used as ID estimators.\\

Although these methods may entail some disadvantages due to the finiteness of datasets \cite{The90}, they have been successfully applied in various fields, such as spatial \cite{Seur09,Chen14} and time series \cite{Dub04} analysis, cosmology \cite{Bor93}, climatology \cite{lov86,lov87,Lov13} and pattern recognition \cite{Hua94,Flo14}. They have also been used in different procedures improving their overall performance \cite{Cam02}. 

\subsection{Maximum Likelihood Estimation Methods}
The Maximum Likelihood Estimation (MLE) of ID was introduced in \cite{Lev04}. The proposed method relies on the assumption that the $k$-nearest neighbors ($k$-NN) of any point $x_i$ of a point set $X_N=\lbrace x_1,\ldots ,x_N\rbrace$ are stemming from a uniform probability density function $f(x_i)$. As a consequence, for a fixed $x_i$, the observations are treated as a homogeneous Poisson process within a small sphere $S_{x_i}(R)$ of radius $R$ centred at $x_i$. On this basis, the inhomogeneous binomial process $\lbrace N(t,x_i),0\leq t\leq R\rbrace$ with\begin{equation}
N(t,x_i)=\sum_{j=1}^N \mathbbm{1}_{\left\lbrace   x_j\in S_{x_i}(t)\right\rbrace}  
\end{equation}counts the number of observations within the distance $t$ of $x_i$ and can be approximated as a Poisson process. The rate of this process is:
\begin{equation}
\lambda(t,x_i)=f(x_i)\ V(m(x_i))\ m(x_i) \ t^{m(x_i)-1}
\end{equation}where $m(x_i)$ is the dimension of the manifold on which $x_i$ lies and $V(m(x_i))$ is the volume of the unit sphere in $\mathbb{R}^{m(x_i)}$ centred at $x_i$. The log-likelihood function of $N(t,x_i)$ can then be expressed as:
\begin{equation}
L(m(x_i),\theta (x_i))=\int_0^R \log(\lambda (t,x_i))\ dN(t,x_i)-\int_0^R \lambda (t,x_i)\ dt
\end{equation}where $\theta(x_i) = \log(f(x_i))$. Finally, the MLE for $m(x_i)$ provides a local estimator of ID \cite{Lev04,Eber07,Ram07}:
\begin{equation}\label{mkx}
\hat{m}_k(x_i) = \left[ \frac{1}{k-2} \sum_{j=1}^{k-1} \log \left(\frac{T_k(x_i)}{T_j(x_i)}\right)\right]^{-1} 
\end{equation}where $k>2$ is the number of NN taken into account and $T_k(x_i)$ is the distance between $x_i$ and its $k^{th}$ NN.\\ 

If it is assumed that all the observations belong to the same manifold, one has that:  
\begin{equation} \label{mk}
\hat{m}_k = \frac{1}{N} \sum_{i=1}^{N}\hat{m}_k(x_i)
\end{equation}which is simply an average over the whole dataset and, for $k \in \lbrace k_1,k_1+1,...,k_2\rbrace$ with $k_1 > 2$, the final estimate of ID is provided in \cite{Lev04}:
\begin{equation} \label{m}
\hat{m} = \frac{1}{k_2-k_1+1} \sum_{k=k_1}^{k_2}\hat{m}_k
\end{equation}A delicate issue which arises from Equations \ref{mkx}, \ref{mk} and \ref{m} is the range of the values $k$ to be chosen. In practical applications, this is similar to the choice of the scale interval in the fractal-based methods. Here, as well, the finiteness of datasets may greatly influence the final estimate of ID if the considered values $k$ are not carefully selected. In \cite{Lev04}, it is advocated to retain a range of small to moderate values, so that each $S_{x_i}(R)$ is small enough to ensure $f(x_i)\approx const$ and large enough to contain sufficiently many points.\\

In \cite{MacKay05}, a modified version of the MLE algorithm is proposed. It consists in averaging the inverse of the $N$ estimators $\hat{m}_k(x_i)$ of Equation \ref{mk}, so that the final estimator of Equation \ref{m} is replaced with:
\begin{equation} \label{mKay}
\hat{m}= \frac{1}{k_2-k_1+1} \sum_{k=k_1}^{k_2}\left[ \frac{1}{N (k-1)} \sum_{i=1}^{N}\sum_{j=1}^{k-1} \log \left(\frac{T_k(x_i)}{T_j(x_i)}\right)\right] ^{-1}
\end{equation} Although the second version of the algorithm is better for small values of $k$, both of them yield similar results \cite{Ram07} and have been successfully applied in various studies \cite{Lev04,MacKay05,Eber07}. Finally, notice that, in the remainder of this paper, the estimators of Equations \ref{m} and \ref{mKay} will be named after their authors, $\hat{m}_{LB}$ and $\hat{m}_{MG}$.

\begin{figure}[H]
\begin{center}
\includegraphics[width=13.5cm]{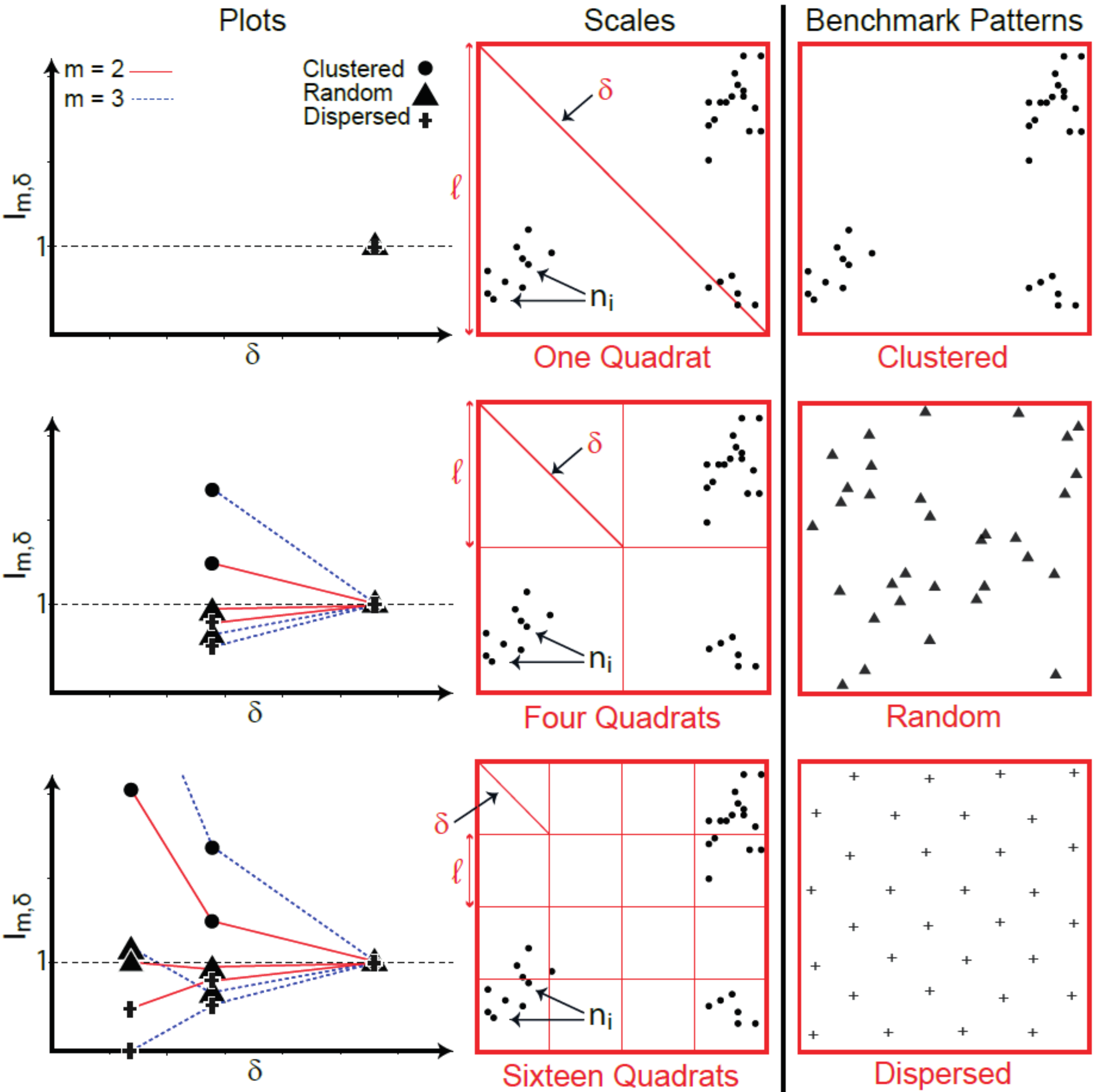}
\end{center}
\caption{Computation of the $m$-Morisita index in two dimensions for $m=2$ and $m=3$ and for three benchmark patterns}
\label{MindexDrawing}
\end{figure}

\section{A New Estimator of Intrinsic Dimension}\label{Mindex}

\subsection{The $m$-Morisita Index}
The $m$-Morisita index \cite{Go14,Hul90}, $I_{m,\delta}$, is a global measure of clustering. It is a generalization of the Morisita index \cite{Mori59,TuiaKan08} and it was first proposed in \cite{Hul90} for the analysis of population distributions in ecology. It was later modified in \cite{Go14} to take into account the notion of scale in spatial data analysis and a relationship to multifractality was established.\\

For its computation, $I_{m,\delta}$, requires a dataset to be covered with a grid of $Q$ quadrats (or cells) of changing size $\delta$ (see Figure \ref{MindexDrawing}). For a fixed $\delta$, $I_{m,\delta}$ measures how many times more likely it is that $m$ ($m \ge 2$) randomly selected data points will be from the same quadrat than it would be if the $N$ points of the dataset were distributed at random (see Figure \ref{MindexDrawing}). It is calculated as follows: \begin{equation}\label{eq:mMorisita}
I_{m,\delta}=Q^{m-1}\frac{\sum_{i=1}^Q n_i(n_i-1)(n_i-2) \dotsm (n_i-m+1)}{N(N-1)(N-2) \dotsm (N-m+1)}
\end{equation}where $n_i$ is the number of points in the $i^{th}$ quadrat and $N$ is the total number of points. The computation of the index starts with a relatively large quadrat size $\delta$. It is then reduced until it reaches a minimum value and a plot relating every $I_{m,\delta}$ to its matching $\delta$ can be drawn.\\

Figure \ref{MindexDrawing} illustrates the computation of the index in two dimensions for three benchmark point distributions (or patterns), for three different scales and for $m=2$ and $m=3$. For the highest possible $\delta$, when only one quadrat is considered, $I_{m,\delta}$ returns the same value for each pattern and $m$. As the number of quadrats increases, $I_{m,\delta}$ adopts a specific behaviour for each of the three benchmark distributions. If the points are distributed at random, every computed $I_{m,\delta}$ oscillates around the value of 1. If the points are clustered, the value of the index increases as $\delta$ decreases and, finally, if the points are dispersed, the index approaches 0 at small scales \cite{Go14, KanMai04}. Further, as $m$ increases, $I_{m,\delta}$ becomes more and more sensitive to the structure of the pattern under study. In complex situations, $I_{m,\delta}$ computed with small $m$ may miss structures which are detected with higher $m$.

\subsection{The Morisita Estimator of Intrinsic Dimension}\label{mID}
Several parallels can be drawn between the $m$-Morisita index and Rényi's information of $q^{th}$ order. In particular, it was established, for fractal point sets, that \cite{Go14}: \begin{equation}  \label{MI_Dq} \lim_{\delta \to 0}\frac{\log{(I_{m,\delta})}}{\log (\frac{1}{\delta})} \frac{1}{m-1}\approx E-D_{m} = C_{m}\end{equation} where $ m \in\lbrace 2,3,4,\dotsm\rbrace$, $C_{m}$ is the codimension of order $q=m$, $E$ is the dimension of the Euclidean space where the dataset is embedded and $D_{m}$ is Rényi's generalized dimension of order $q=m$. In practical applications, for finite datasets, it can be shown that Equation \ref{MI_Dq} is verified only under the condition that $H:=\underset{i}{max}\left(n_i\right) \gg m$ \cite{Go14}. If so, $C_m$ can be estimated from the slope, $S_{m}$, of the straight line fitting the linear part of the plot relating $\log{(I_{m,\delta})}$ to $\log \left( \frac{1}{\delta}\right) $. Then, $C_{m}\approx S_{m}/(m-1)$ and one has that:
\begin{equation} \label{est_mindex_1}
D_m \approx E - \left( \frac{S_{m}}{m-1}\right) 
\end{equation}where $ m \in\lbrace 2,3,4,\dotsm\rbrace$.\\

In high-dimensional spaces, the condition $H\gg m$ is hardly ever met at small scales. In such situations, it is important to notice that the major difference between $\log{(I_{m,\delta})}$ and $RI_{m}(\delta)$ lies in the following inequality:\begin{equation}  \label{MI_Dq_ineq} n_i^m > n_i(n_i-1) \dotsm (n_i-m+1) \end{equation}Consequently, unlike $\log{(I_{m,\delta})}$, $RI_{m}(\delta)$ seriously overestimates the probability of randomly drawing $m$-tuples of points from grid cells characterized by a small $n_i$. Such cells are numerous at small scales or when the sample size is limited and can greatly affect the accuracy of $D_m$. From this perspective, it is possible to suggest a new ID estimator based on $\log{(I_{m,\delta})}$ (for $m\geq 2$):\begin{equation} \label{est_mindex_2}
M_m := E - \left( \frac{S_{m}}{m-1}\right) 
\end{equation}which should be more robust to sample size than $D_m$ (see Equation \ref{Dq_cla}). Notice that $M_m$ should only be computed under the condition that $H>m$ at all considered scales. In the remainder of this paper, $M_m$ will be referred to as the Morisita estimator of ID. It will be thoroughly tested in Section \ref{appcomp} with synthetic datasets of various complexities and with real data from the UCI machine learning repository.

\begin{algorithm}[t]
\caption{The Morisita INDex for ID estimation (MINDID)}\label{mindex_algo}
\textbf{INPUT:} a $N\times E$ matrix, $D$, with $N$ points and $E$ features; a vector $L$ of values $\ell$; a vector $M$ of contiguous values $m$. 
\textbf{OUTPUT:} a vector containing $M_m$ for each value $m$. 
\begin{algorithmic}[1]
\STATE Rescale each variable to $[0,1]$
\FORALL{values $\ell$}
\STATE Divide each element of $D$ by $\ell$ and round the result to the next lowest whole number in a $N\times E$ matrix called $D\ell$
\STATE Count the number, $nbr\_l$, of different lines in $D\ell$ and store their frequency in a vector, $ni$, of size $nbr\_l$
\FORALL{values $m$}
\STATE{Compute $\log{(I_{m,\delta})}$ (See Equation \ref{eq:log_mMorisita}) using the vector $ni$ and store the result in a $\mid L\mid \times \mid M\mid$ matrix called logMindex}
\ENDFOR
\ENDFOR\\
\STATE Optional: Compute the values $\delta$ (See Equation \ref{ldelta}) using the vector $L$ and store the result in a vector $\Delta$ of cardinality $\mid L\mid$
\FORALL{values $m$}
\STATE Compute $M_m$ (See Equation \ref{est_mindex_2}) using logMindex and $\Delta$ (or $L$)
\ENDFOR
\end{algorithmic}
\end{algorithm} 

\subsection{An Algorithm for Large Datasets}
Algorithms used for handling large datasets should be affected as little as possible by the amount of main memory available. Regarding $M_m$, the main issue concerns the way the number of points per quadrat is counted. A good algorithm must be able to effectively disregard empty quadrats. It is also more appealing if its implementation is straightforward in most programming environments (e.g. R and Matlab). To fulfil these requisites, the MINDID algorithm (see Algorithm \ref{mindex_algo}) incorporates a version of an algorithm suggested in \cite{Zha02}. It rescales each variable of a dataset $D$ to $[0,1]$ and takes advantage of the properties of the square cells of the hyper-grid covering the data: for a given cell size $\delta$, each value of the dataset is divided by $\ell$ (see Figure \ref{MindexDrawing}) and rounded to the next lowest integer. In this way, in the resulting matrix, all the data points falling into the same cell are matched by as many equal lines.\\ 

Another issue concerns $Q^{m-1}$, since it is often given the value \textit{Inf} (e.g. R and Matlab) for small values of $\ell$ when $E\gg 1$. A way to overcome this problem is to resort to $\log{(I_{m,\delta})}$ instead of $I_{m,\delta}$. $Q^{m-1}$ is related to $\ell$ and $E$ through:\begin{equation} Q^{m-1} = \left( \frac{1}{\ell}\right) ^{E(m-1)}\end{equation} and \begin{equation} \label{eq:log_mMorisita}\log{(I_{m,\delta})} = E(1-m)\log{\left(\ell\right) }+\log{\left( \frac{\sum_{i=1}^Q n_i(n_i-1) \dotsm (n_i-m+1)}{N(N-1) \dotsm (N-m+1)}\right)}\end{equation}This solution is satisfactory, since the computation of $M_m$ (See Equation \ref{est_mindex_2}) only requires $\log{(I_{m,\delta})}$. Consequently, the second part of the MINDID algorithm is devoted to the implementation of Equation \ref{eq:log_mMorisita}.\\

Finally, the computation of $M_m$ is carried out using either $\log(\ell)$ or $log(\delta)$. In the rest of this paper, the second option will be preferred. Notice that $\ell$ and $\delta$ are related as follows:\begin{equation} \label{ldelta} \delta = \ell\sqrt{E}\end{equation}

\section{Assessment of the Morisita Estimator of ID}\label{appcomp}

\subsection{Synthetic Data}\label{sec_data}
\begin{figure}[H]
\begin{center}
\includegraphics[width=13.5cm]{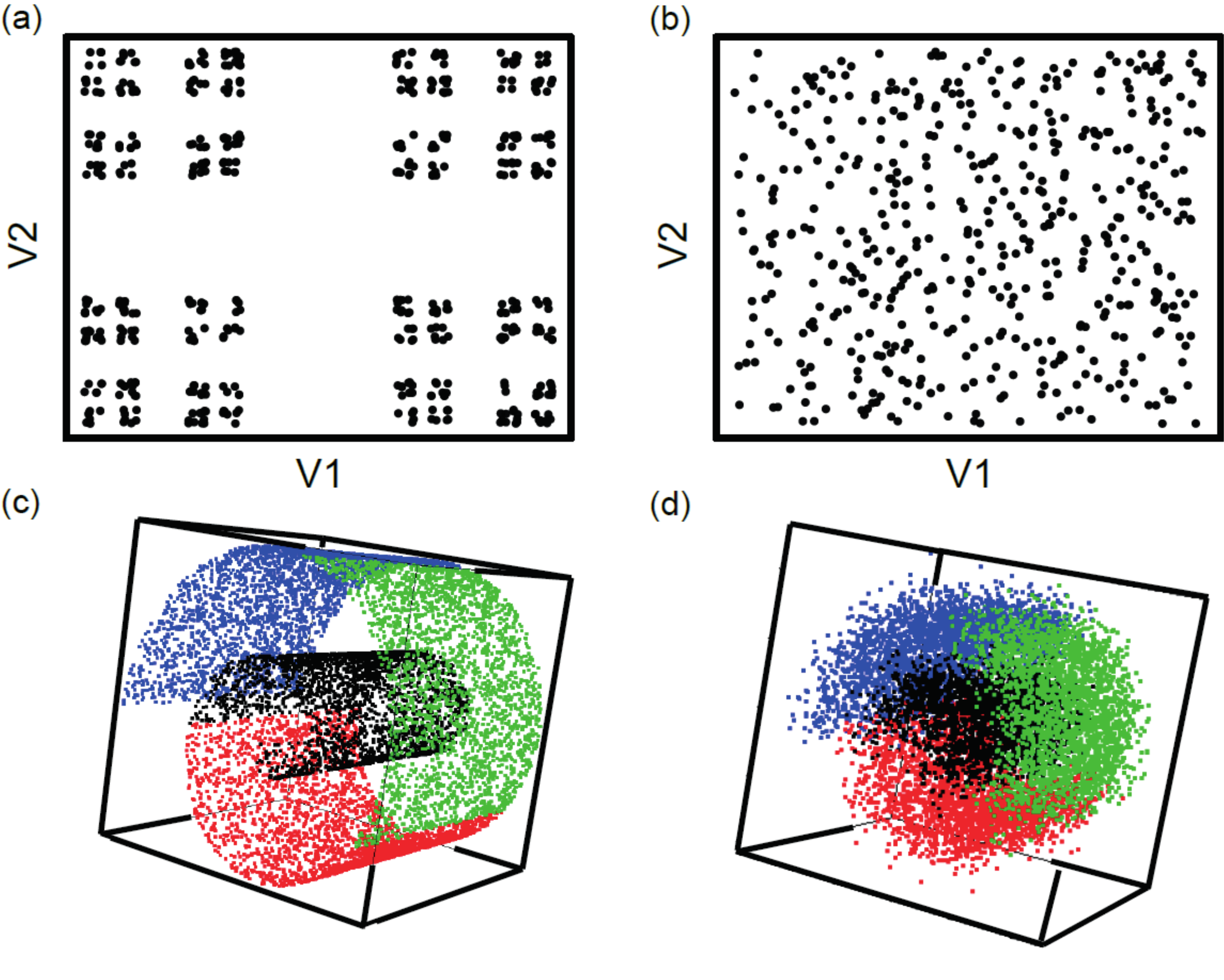}
\end{center}
\caption{(a) A Cantor set of $512$ points for $E=2$; (b) a realization of a uniform cloud of $500$ points for $E=2$; (c) a Swiss roll of $10 \, 000$ points without noise and (d) a Swiss roll of $10 \, 000$ points with noise $G \sim N(0,0.25^2)$}
\label{data3}
\end{figure}

Several datasets were built, so that each of them resides on a known manifold (or near a known manifold in the case of noisy data). They can be divided into four categories (see Figure \ref{data3}):
\begin{enumerate}
\item Swiss rolls (e.g. \cite{LeeVer07}) of $1000$, $5000$ and $10 \, 000$ points. The theoretical ID of the data is equal to 2.
\item Noisy Swiss rolls of $1000$, $5000$ and $10 \, 000$ points. The noise is modelled as a Gaussian variable $G \sim N(0,\sigma^2)$ where $\sigma$ varies from $0$ to $0.5$.
\item Uniform clouds of $1000$, $5000$ and $10 \, 000$ points. Each of the $N$ distinct points $x_i$, with $i \in \lbrace 1,2,\cdots, N\rbrace$, is described by a E-dimensional vector $[x_i^1,x_i^2,$ $\cdots,x_i^E]^T \in \mathbb{R}^E$, the components of which are sampled from $E$ i.i.d. variables following a uniform distribution. $E$ is gradually increased from $1$ to $7$ and is equal to the theoretical ID of the data. 
\item This last category is based on the properties of the Cartesian product of some fractals \cite{Fal03,Shi95}. One-dimensional Cantor sets of $8192$ and $65 \, 536$ points are first created. The resulting vectors are shuffled seven times to generate as many variables. A Euclidean space $\mathbb{R}^E$ can then be constructed and $E$ is gradually increased from 1 to 7. The dimension of the data manifold (i.e. the theoretical ID) is equal to $\frac{\log(2)}{\log(3)}E$, where $\frac{\log(2)}{\log(3)}$ is the Hausdorff dimension of a one-dimensional Cantor set.
\end{enumerate}

\subsection{Comparison Between $M_m$ and $D_m$ Using Synthetic Data}
\begin{figure}[H]
\begin{center}
\includegraphics[width=13.5cm]{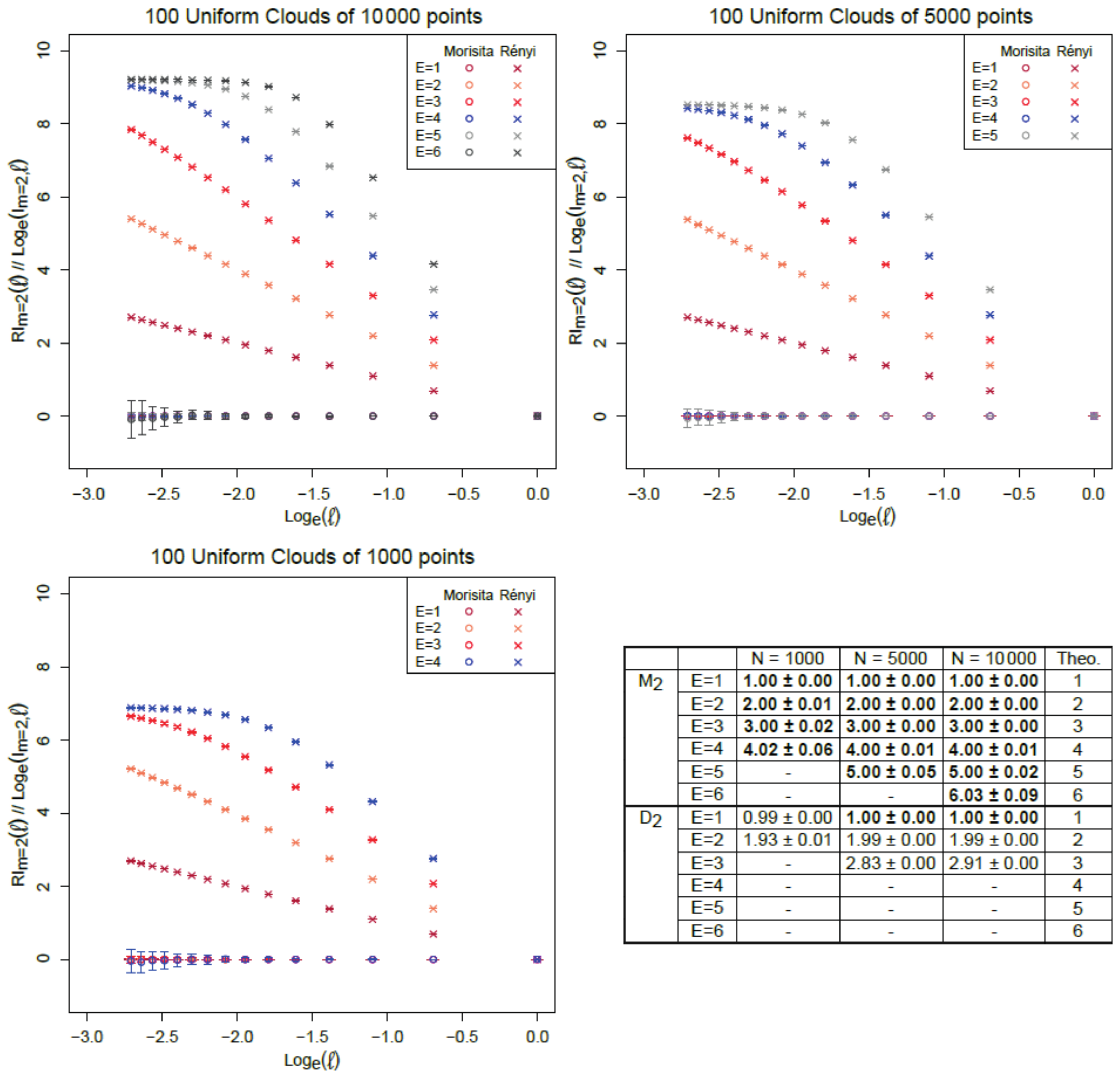}
\end{center}
\caption{The results of the application of $I_{2,\ell}$ and $RI_{2}(\ell)$ to 100 uniform point clouds of $10 \,000$, $5000$ and $1000$ points. In the bottom-right table, the corresponding estimates of $D_2$ and $M_2$ are provided as follows: the mean (computed over the 100 sets) $\pm$ the standard deviation. The results written in bold script indicate that the theoretical ID falls within the mean value $\pm$ one standard deviation.}
\label{rnd_Cloud}
\end{figure}

\begin{figure}[H]
\begin{center}
\includegraphics[width=13cm]{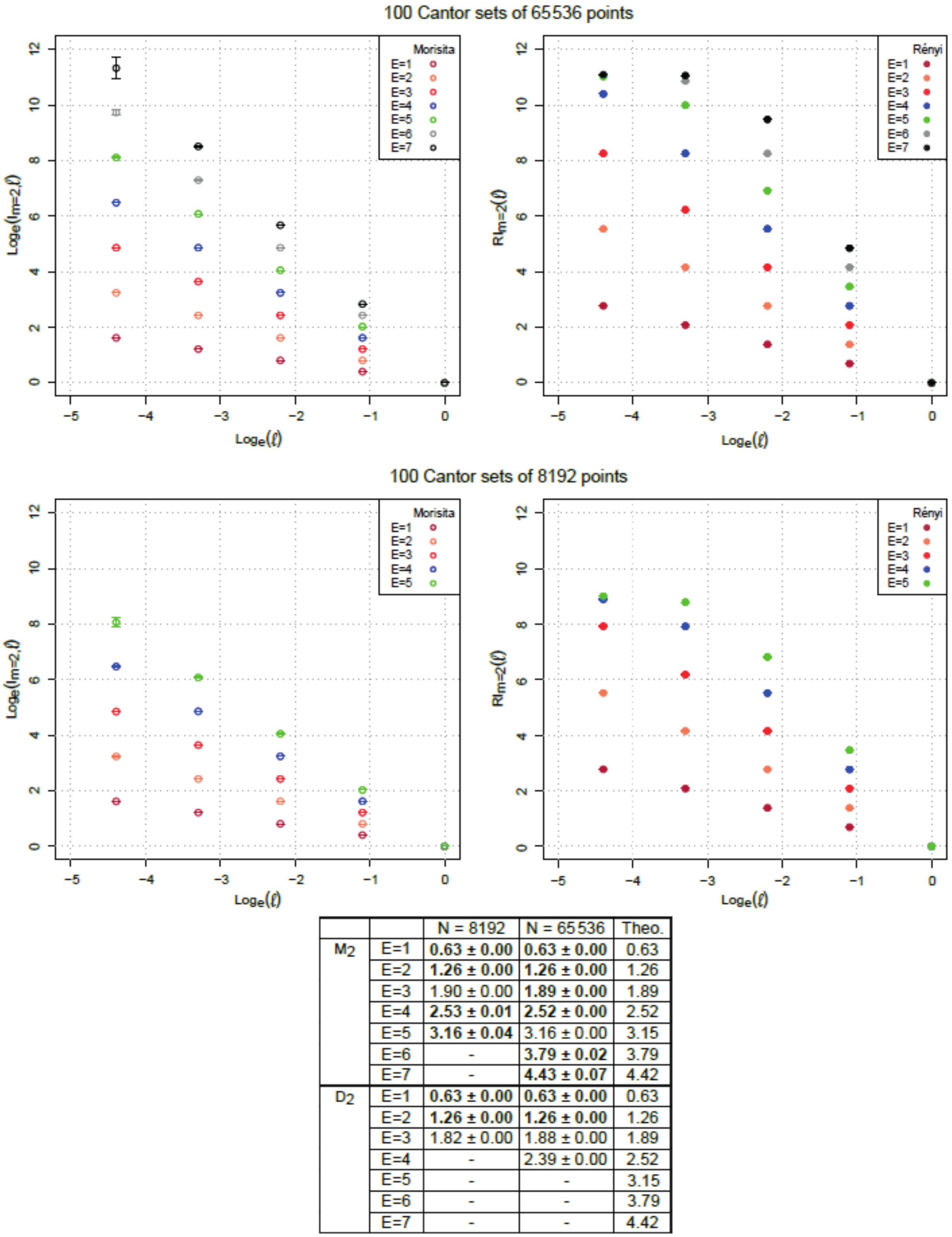}
\end{center}
\caption{The results of the application of $I_{2,\ell}$ (left) and $RI_{2}(\ell)$ (right) to 100 Cantor sets of $8192$ and $65 \, 536$ points. In the bottom table, the corresponding estimates of $D_2$ and $M_2$ are provided as follows: the mean (computed over the 100 sets) $\pm$ the standard deviation. The results written in bold script indicate that the theoretical ID falls within the mean value $\pm$ one standard deviation.}
\label{Cantor_set}
\end{figure}

Two categories of datasets were used to compare $M_m$ with $D_m$: the uniform point clouds and the Cantor sets. They were employed with varying $N$ and $E$ (see Subsection \ref{sec_data}) and for each combination of these two parameters, 100 sets were generated. The results are displayed in Figures \ref{rnd_Cloud} and \ref{Cantor_set} and in Tables \ref{N1} and \ref{N2}.\\

In the case of the uniform point clouds, $RI_2(\ell)$ and $\log\left(I_{2,\ell}\right)$ were computed by using an interval of the parameter $\ell^{-1}$ ranging from 1 to 15 (see Equation \ref{ldelta}). Figure \ref{rnd_Cloud} shows the results for $N= 10 \, 000$, $N=5000$ and $N=1000$ and for increasing $E$. The points of each plot are the mean values yielded by the two indices over the 100 sets and the error bars correspond to the standard deviations.

If the value of $H$ (see Subsection \ref{mID}) was smaller than 2 at certain scales $\ell$ for a given dimension $E$, the entire plots describing the corresponding behaviours of $RI_2(\ell)$ and $\log\left(I_{2,\ell}\right)$ were not drawn. This was motivated by the condition that $H$ must be greater than $m$ at all scales for $M_m$ to be computed.\\

In spite of this limitation, $M_m$ provides better ID estimates than $D_q$ when the same values of $\ell$ are considered. In Figure \ref{rnd_Cloud}, this is highlighted, for the largest $E$, by the steady state reached by $RI_2(\ell)$ at small scales, which shows a departure from the power law of Equation \ref{renyione}. As a consequence, $D_2$ cannot be derived from a linear regression calculated over the whole range of $\ell$. In contrast, $\log\left(I_{2,\ell}\right)$ follows, on average and throughout the scales, the power law which underlies Equation \ref{est_mindex_2}: all the plots are superimposed on a constant mean level of 0 as expected from a Poisson distribution (i.e. a random space-filling set). The cost of this near absence of bias is an increase in the variability of the values provided by $\log\left(I_{2,\ell}\right)$ as $\ell$ decreases. As indicated in the table of Figure \ref{rnd_Cloud}, this drawback has only a small impact on the variabilities of the final ID estimates which remain low.\\ 

Still with regard to the table of Figure \ref{rnd_Cloud}, the means and standard deviations of the ID estimates\footnote{The linear slope of the plots must be multiplied by $-1$ to yield $S_m$ and $D_q$, since the x-axis represents $\ell$ instead of $\ell^{-1}$} were calculated only if the dependence between $\log\left(I_{2,\ell}\right)$ or $RI_2(\ell)$ and $\ell$ could be reasonably approximated by using a linear regression over all the scales (i.e if only one slope could be distinguished in the different plots). The results show that $D_2$ becomes unreliable for $E > 2$, while $M_2$ works better and can even be used up to $E=6$ for $N= 10 \, 000$.\\

The results of $M_2$ and $D_2$ were written in bold script if the theoretical ID fell within the mean value $\pm$ one standard deviation. Shapiro-Wilk tests were first conducted to check whether or not the estimates could be assumed to come from normal distributions (only if the standard deviations were not equal to $0.00$). At a $5\%$ $\alpha$ level, the hypothesis of normality could not be rejected for any of the estimate distributions. The theoretical ID fell within the mean value $\pm$ one standard deviation for each result yielded by $M_2$. Regarding $D_2$, it only happened twice for high values of $N$ and for $E=1$. Notice that the same notation will be used throughout the section.    

Finally, the number of $E$-dimensional spaces, for which $M_2$ can be calculated, decreases as $N$ is reduced. The reason is that it becomes less likely that at least two points will fall into the same cell at small scales when $E \gg 1$. Nevertheless, whatever $N$, the bias affecting $RI_2(\ell)$ is always noticeable for the greatest $E$ and tends to lead to an underestimation of ID in any case.\\

Similar comments can be made about the results obtained for the Cantor sets (see Figure \ref{Cantor_set}). For this second category of data, the interval of the parameter $\ell^{-1}$ follows a geometric series with ratio $r=3$ and ranges from 1 to 81. In this way, the grid used for the computation of both $\log\left(I_{2,\ell}\right)$ and $RI_2(\ell)$ is in accordance with the mathematical construction of a Cantor set. For the same arguments as those previously set out, $M_2$ turns out to be a more reliable estimator of ID than $D_2$: here as well, a bias affects the behaviour of $RI_2(\ell)$ when the number of points in occupied cells is low. It is also interesting to notice that $M_2$ can be computed up to $E=5$ for $N=8192$, although the lowest considered scale $\ell$ is smaller than it was for the uniform point clouds. This is due to the dimension of the data manifold that is systematically less than $E$. As a consequence, a Cantor set is not space-filling and it is more likely that at least two points will fall into the same cell than it would be if the $N$ data points were randomly distributed within the entire Euclidean space. This observation highlights that the curse of ID \cite{Trai10} (i.e. the problems induced by a high ID) is a central issue when studying high-dimensional spaces.\\

\begin{table}
\centering
\tiny
\renewcommand{\arraystretch}{1.2}
\begin{tabular}{|c||c|c|c|c|c|c|}
\hline 
$N = 65 \, 536$ & $M_2$ & $D_2$ & $M_3$ & $D_3$ & $M_5$ & $D_5$\\ 
\hline
\hline 
$E=1$ & \mbox{\boldmath${0.63 \pm 0.00}$} & \mbox{\boldmath${0.63 \pm 0.00}$} & \mbox{\boldmath${0.63 \pm 0.00}$} & \mbox{\boldmath${0.63 \pm 0.00}$} & \mbox{\boldmath${0.63 \pm 0.00}$} & \mbox{\boldmath${0.63 \pm 0.00}$}\\ 
\hline 
$E=2$ & \mbox{\boldmath${1.26 \pm 0.00}$} & \mbox{\boldmath${1.26 \pm 0.00}$} & \mbox{\boldmath${1.26 \pm 0.00}$} & $1.26 \pm 0.00$ & \mbox{\boldmath${1.26 \pm 0.00}$} & \mbox{\boldmath${1.26 \pm 0.00}$}\\ 
\hline 
$E=3$ & \mbox{\boldmath${1.89 \pm 0.00}$} & $1.88 \pm 0.00$ & \mbox{\boldmath${1.89 \pm 0.00}$} & $1.88 \pm 0.00$ & \mbox{\boldmath${1.89 \pm 0.00}$} & $1.87 \pm 0.00$\\ 
\hline 
$E=4$ & \mbox{\boldmath${2.52 \pm 0.00}$} & $2.39 \pm 0.00$ & \mbox{\boldmath${2.52 \pm 0.00}$} & $2.37 \pm 0.00$ & \mbox{\boldmath${2.53 \pm 0.01}$} & $2.33 \pm 0.00$\\ 
\hline 
$E=5$ & $3.16 \pm 0.00$ & - & \mbox{\boldmath${3.16 \pm 0.02}$} & - & - & -\\ 
\hline 
$E=6$ & \mbox{\boldmath${3.79 \pm 0.02}$} & - & - & - & - & - \\ 
\hline 
$E=7$ & \mbox{\boldmath${4.43 \pm 0.07}$} & - & - & - & - & -\\ 
\hline
\end{tabular}
\caption{The results of the application of $M_m$ and $D_m$ to 100 Cantor sets of $65 \, 536$ points for $m=2$, $m=3$ and $m=5$. The mean $\pm$ the standard deviation is provided for each $E$-dimensional space. The results written in bold script indicate that the theoretical ID falls within the mean value $\pm$ one standard deviation.}
\label{N1}
\end{table}

\begin{table}
\centering
\tiny 
\renewcommand{\arraystretch}{1.2}
\begin{tabular}{|c||c|c|c|c|c|c|}
\hline 
$N = 8192$ & $M_2$ & $D_2$ & $M_3$ & $D_3$ & $M_5$ & $D_5$ \\ 
\hline
\hline 
$E=1$ & \mbox{\boldmath${0.63 \pm 0.00}$} & \mbox{\boldmath${0.63 \pm 0.00}$} & \mbox{\boldmath${0.63 \pm 0.00}$} & \mbox{\boldmath${0.63 \pm 0.00}$} & \mbox{\boldmath${0.63 \pm 0.00}$} & \mbox{\boldmath${0.63 \pm 0.00}$} \\ 
\hline 
$E=2$ & \mbox{\boldmath${1.26 \pm 0.00}$} & \mbox{\boldmath${1.26 \pm 0.00}$} & \mbox{\boldmath${1.26 \pm 0.00}$} & $1.25 \pm 0.00$ & \mbox{\boldmath${1.26 \pm 0.00}$} & $1.25 \pm 0.00$ \\ 
\hline 
$E=3$ & $1.90 \pm 0.00$ & $1.82 \pm 0.00$ & $1.90 \pm 0.00$ & $1.80 \pm 0.00$ & $1.90 \pm 0.01$ & $1.76 \pm 0.00$ \\ 
\hline 
$E=4$ & \mbox{\boldmath${2.53 \pm 0.01}$} & - & \mbox{\boldmath${2.53 \pm 0.03}$} & - & - & - \\ 
\hline 
$E=5$ & \mbox{\boldmath${3.16 \pm 0.04}$} & - & - & - & - & - \\ 
\hline 
$E=6$ & - & - & - & - & - & -  \\ 
\hline 
$E=7$ & - & - & - & - & - & - \\ 
\hline
\end{tabular}
\caption{The results of the application of $M_m$ and $D_m$ to 100 Cantor sets of 8192 points for $m=2$, $m=3$ and $m=5$. The mean $\pm$ the standard deviation is provided for each $E$-dimensional space. The results written in bold script indicate that the theoretical ID falls within the mean value $\pm$ one standard deviation.}
\label{N2}
\end{table}

The Cantor sets were also used to assess the accuracy of $M_m$ and $D_m$ for $m$ greater than 2. Tables \ref{N1} and \ref{N2} show the results for, respectively, $N= 65 \, 536$ and $N=8192$. In both cases, $M_m$ provides results closer to the theoretical ID than those yielded by $D_m$ for $E>2$. Nevertheless, the difference between the two estimators is lessened as the value of $m$ is increased. This follows from Equation \ref{MI_Dq_ineq}: when $m$ ($m=q$) is high, the impact of the inequality is negligible, while the implementation of $M_m$ requires more data points than that of $D_m$.

\subsection{Comparison Between $M_2$ and the Distance-Based Estimators Using Synthetic Data}
$M_2$ was also compared to the other estimators of ID presented in Section \ref{IDest}, namely $d_{cor}$, $\hat{m}_{LB}$ and $\hat{m}_{MG}$. The Swiss rolls and the uniform point clouds were used for this task. The parameters of each estimator were set using the Swiss rolls of 1000 points and stayed unchanged in the whole subsection. At each step, it was made sure that these parameters were close to the ideal ones. Regarding $M_2$, the interval of the parameter $\ell^{-1}$ was chosen, so that it ranged from 5 to 15 and the two MLE estimators were computed with $k$ going from 10 to 20. More challenging, the interval of the parameter $\delta$ of $d_{cor}$ turned out to be relatively complicated to set, since it tended to deviate from the ideal values as both $N$ and $E$ were increased. It was finally decided to resort to percentiles and the problem was empirically solved as follows: (a) 1 percent of the pairwise distances between points had to be lower than the smallest $\delta$ and 7 percent of them had to be lower than the largest one; (b) the range of the interval was divided by 100 to produce intermediate values of $\delta$. It is also worth mentioning that each variable was rescaled, so that it ranged from 0 to 1. Such a transformation is mandatory when working with data of different nature. Regarding $d_{cor}$, $\hat{m}_{LB}$ and $\hat{m}_{MG}$, it amounts to using the Mahalanobis distance and, if the data are not noisy, it has, of course, no influence on $M_m$ (or $D_q$), since the grid employed in its computation is also transformed.\\

\begin{table}
\centering 
\begin{tabular}{|c||c|c|c|}
\hline 
Swiss Roll & $N=1000$ & $N=5000$ & $N = 10 \, 000$\\ 
\hline 
\hline 
$M_2$ &  \mbox{\boldmath${2.03\pm 0.03}$} & $2.03\pm 0.01$ & $2.03 \pm 0.00$\\ 
\hline 
$d_{cor}$ &  $1.95\pm 0.01$ & $1.94\pm 0.00$ & $1.94 \pm 0.00$\\ 
\hline 
$\hat{m}_{LB}$ & $1.95\pm 0.02$ & $1.98\pm 0.01$ & $1.98 \pm 0.01$\\ 
\hline 
$\hat{m}_{MG}$ & $1.95\pm  0.02$ & $1.98\pm 0.01$ & $1.98 \pm 0.00$\\ 
\hline 
\end{tabular}
\caption{The results of the application of the four estimators to Swiss rolls of $N=1000$, $N=5000$ and $N=10 \, 000$ points (100 sets for each $N$). The mean $\pm$ the standard deviation of the estimators, computed over the 100 sets, is provided for each $N$. The result written in bold script indicates that the theoretical ID falls within the mean value $\pm$ one standard deviation (the theoretical ID of the Swiss Roll is 2).}
\label{N3}
\end{table}

\begin{table}
\centering
\footnotesize  
\begin{tabular}{|c||c|c|c|c|c|}
\hline 
 &  & $M_2$ & $d_{cor}$ & $\hat{m}_{LB}$ & $\hat{m}_{MG}$ \\ 
\hline 
\hline 
\multirow{7}{*}{$N=1000$} & $E=1$ & \mbox{\boldmath${1.00\pm 0.00}$} &  \mbox{\boldmath${0.99 \pm 0.01}$} &  \mbox{\boldmath${1.00 \pm 0.01}$} &  \mbox{\boldmath${1.00 \pm 0.01}$} \\ 
\cline{2-6} 
           & $E=2$ & \mbox{\boldmath${2.00\pm 0.01}$} & $1.91 \pm 0.01$ & $1.95 \pm 0.02$ & $1.95 \pm 0.02$ \\ 
\cline{2-6} 
           & $E=3$ & \mbox{\boldmath${3.00\pm 0.05}$} & $2.74 \pm 0.02$ & $2.83 \pm 0.02$ & $2.83 \pm 0.03$ \\ 
\cline{2-6} 
           & $E=4$ & \mbox{\boldmath${4.04 \pm 0.19}$} & $3.51 \pm 0.03$ & $3.66 \pm 0.03$ & $3.65 \pm 0.04$ \\ 
\cline{2-6}
           & $E=5$ & - & $4.24 \pm 0.04$ & $4.45 \pm 0.05$ & $4.44 \pm 0.05$ \\ 
\cline{2-6}
           & $E=6$ & - & $4.93 \pm 0.04$ & $5.20 \pm 0.06$ & $5.20 \pm 0.06$ \\ 
\cline{2-6}
           & $E=7$ & - & $5.59 \pm 0.04$ & $5.92 \pm 0.07$ & $5.92 \pm 0.07$ \\ 
\hline
\hline  
\multirow{7}{*}{$N=5000$} & $E=1$ & \mbox{\boldmath${1.00\pm 0.00}$} & \specialcell{$0.99\pm 0.00$\\\ \mbox{\boldmath${\mathit{1.00\pm 0.00}}$}} & \mbox{\boldmath${1.00\pm 0.00}$} & \mbox{\boldmath${1.00\pm 0.00}$} \\ 
\cline{2-6} 
           & $E=2$ & \mbox{\boldmath${2.00\pm 0.00}$} & \specialcell{$1.91\pm 0.01$ \\\ $\mathit{1.95\pm 0.00}$} & $1.98\pm 0.01$ & $1.98\pm 0.01$ \\ 
\cline{2-6}  
           & $E=3$ & \mbox{\boldmath${3.00\pm 0.01}$} & \specialcell{$2.74\pm 0.01$\\\ $\mathit{2.84\pm 0.01}$} & $2.90\pm 0.01$ & $2.90\pm 0.01$ \\ 
\cline{2-6}  
           & $E=4$ & \mbox{\boldmath${4.00\pm 0.04}$} & \specialcell{$3.51\pm 0.01$\\\ $\mathit{3.68\pm 0.01}$} & $3.78\pm 0.02$ & $3.77\pm 0.02$ \\ 
\cline{2-6}  
           & $E=5$ & \mbox{\boldmath${5.02\pm 0.13}$} & \specialcell{$4.24\pm 0.01$\\\ $\mathit{4.47\pm 0.02}$} & $4.61\pm 0.02$ & $4.60\pm 0.02$ \\ 
\cline{2-6}  
           & $E=6$ & - & \specialcell{$4.93\pm 0.01$\\\ $\mathit{5.22\pm 0.01}$} & $5.40\pm 0.03$ & $5.40\pm 0.03$ \\ 
\cline{2-6}  
           & $E=7$ & - & \specialcell{$5.58\pm 0.01$\\\ $\mathit{5.94\pm 0.01}$} & $6.18\pm 0.03$ & $6.17\pm 0.03$ \\ 
\hline 
\hline
\multirow{7}{*}{$N=10 \, 000$} & $E=1$ & \mbox{\boldmath${1.00\pm 0.00}$} & \specialcell{$0.99\pm 0.00$\\\ \mbox{\boldmath${\mathit{1.00\pm 0.00}}$}} & \mbox{\boldmath${1.00\pm 0.00}$} & \mbox{\boldmath${1.00\pm 0.00}$}\\ 
\cline{2-6} 
           & $E=2$ & \mbox{\boldmath${2.00 \pm 0.00}$} & \specialcell{$1.91\pm 0.00$\\\ $\mathit{1.97\pm 0.00}$}  & $1.98\pm 0.00$ & $1.98\pm 0.01$ \\ 
\cline{2-6} 
           & $E=3$ & \mbox{\boldmath${3.00 \pm 0.00}$} & \specialcell{$2.74\pm 0.01$\\\ $\mathit{2.90\pm 0.01}$}  & $2.93\pm 0.01$ & $2.92\pm 0.01$ \\ 
\cline{2-6} 
           & $E=4$ & \mbox{\boldmath${4.00 \pm 0.02}$} & \specialcell{$3.51\pm 0.01$\\\ $\mathit{3.78\pm 0.01}$}  & $3.82 \pm 0.01$ & $3.81 \pm 0.01$ \\ 
\cline{2-6}  
           & $E=5$ & \mbox{\boldmath${5.00 \pm 0.08}$} & \specialcell{$4.24 \pm 0.01$\\\ $\mathit{4.62 \pm 0.01}$}  & $4.66 \pm 0.02$ & $4.66 \pm 0.02$ \\ 
\cline{2-6}  
           & $E=6$ & \mbox{\boldmath${6.06 \pm 0.24}$} & \specialcell{$4.93 \pm 0.01$\\\ $\mathit{5.42 \pm 0.01}$}  & $5.48 \pm 0.02$ & $5.47 \pm 0.02$ \\ 
\cline{2-6} 
           & $E=7$ & - & \specialcell{$5.58 \pm 0.01$\\\ $\mathit{6.19 \pm 0.02}$}  & $6.27 \pm 0.02$ & $6.26 \pm 0.02$ \\ 
\hline 
\end{tabular} 
\caption{The results of the application of the four estimators to uniform distribution clouds of $N=1000$, $N=5000$ and $N=10 \, 000$ points (100 sets for each $N$). The mean $\pm$ standard deviation, computed over the 100 sets, is provided for each $N$. The results written in bold script indicate that the theoretical ID falls within the mean value $\pm$ one standard deviation (the theoretical ID of a uniform point cloud  is $E$).}
\label{N4}
\end{table}

Tables \ref{N3} and \ref{N4} show the results (mean $\pm$ standard deviation computed over 100 sets) provided by the four estimators for, respectively, the Swiss rolls and the uniform point clouds. The values yielded by the four estimators are similar when the data points are densely distributed on their manifold. This situation is encountered for the Swiss rolls and for relatively low $E$ and high $N$ in the case of the uniform distributions. When the data points are sparse (low $N$ and/or high theoretical ID) and provided that it can be calculated, $M_2$ provides better ID estimates. $d_{cor}$, $\hat{m}_{LB}$ and $\hat{m}_{MG}$ are always able to yield a result, but, when the data are simply too sparse for $M_2$ to be computed, they tend to seriously underestimate the true ID of the data. The underestimation reduces as $N$ increases and it can be assumed that it is caused (at least partially) by edge effects.\\ 

Several Edge effect corrections have been proposed and thoroughly studied in spatial data analysis \cite{Ripl81}, but the problem has often been overlooked in ID estimation methods. $d_{cor}$ appears to be more affected by this problem than the other estimators. Nevertheless, the case of $d_{cor}$ is difficult to deal with, since the ideal range of the parameter $\delta$ is very sensitive to $N$ and $E$. Consequently, concerning the uniform distribution clouds, it was decided that a new series of 10 estimations would be carried out for $N=5000$ and $N= 10 \, 000$ with improved parameters. The results appear in italics in Table \ref{N4} and reveal that the modifications have improved the estimates, even if $\hat{m}_{LB}$ and $\hat{m}_{MG}$ stay better.

\begin{figure}[H]
\begin{center}
\includegraphics[width=13.5cm]{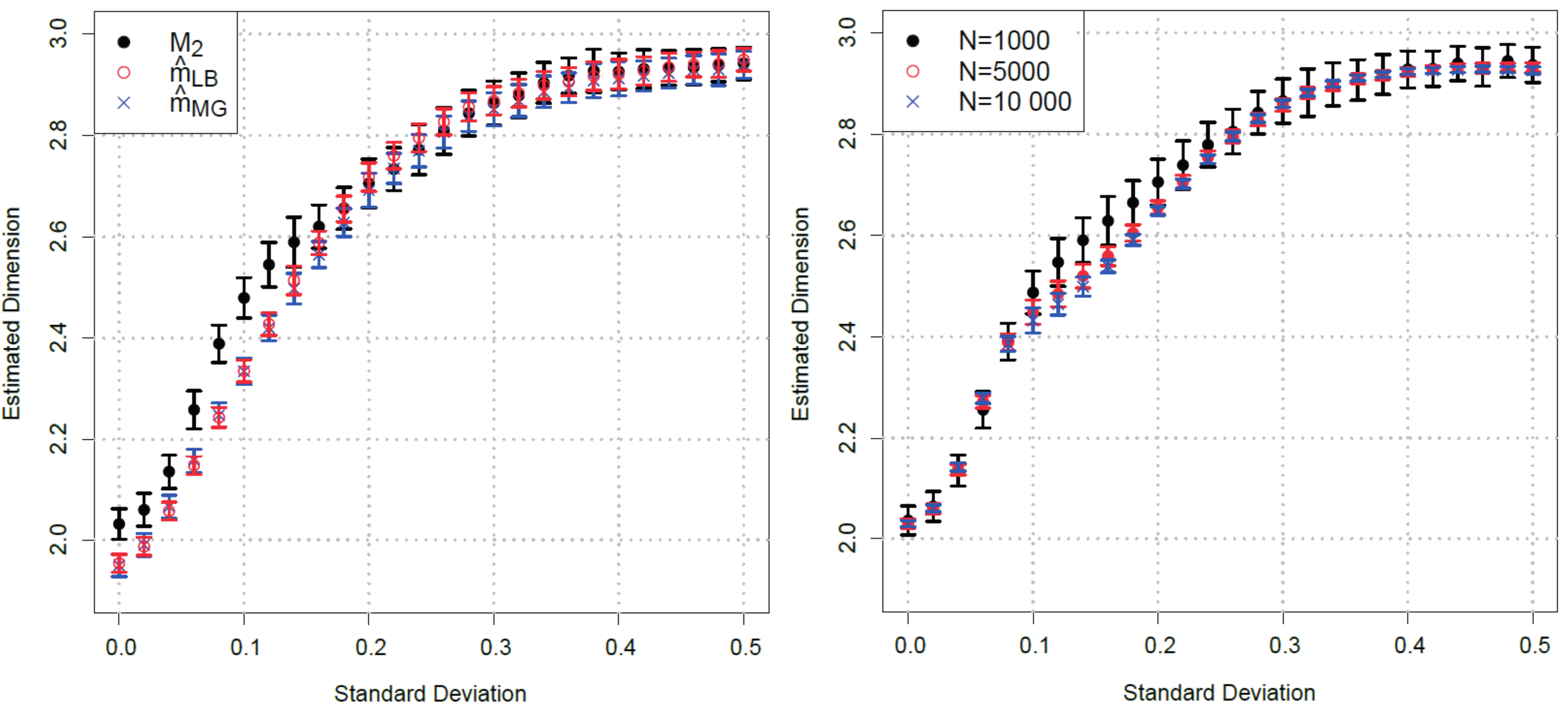}
\end{center}
\caption{(left) Application of $M_2$, $\hat{m}_{LB}$ and $\hat{m}_{MG}$ to noisy Swiss rolls of $N=1000$; (right) Application of $M_2$ to noisy Swiss rolls of $N=1000$, $N=5000$ and $N= 10 \, 000$. The x-axis represents the standard deviation of the noise and, for each level, 100 sets are considered.}
\label{mindex_noise}
\end{figure}

\subsection{Comparison Between the Estimators Using Noisy Synthetic Data}
The presence of noise implies that the data points are located near a manifold instead of being exactly on it \cite{Lev04}. Consequently, a robust ID estimator should be as insensitive to noise as possible. In order to test the robustness of $M_2$, the noisy Swiss rolls, presented in Subsection \ref{sec_data}, were used and the results are displayed in Figure \ref{mindex_noise}. On the left, a comparison between $M_2$, $\hat{m}_{LB}$ and $\hat{m}_{MG}$ was conducted to show how the mean and the standard deviation of the estimates (computed over 100 sets) change as the noise increases. Leaving aside the initial difference, the sensitivity of the three estimators appears to be similar. The right hand-side of the figure highlights that the number of points $N$ has a low influence on the responsiveness of $M_2$ to noise. These results demonstrate that the behaviour of $M_2$ in presence of noise is not better or worse than that of the other two estimators.

\subsection{Comparison Between the Estimators Using Real Data}
$M_2$, $D_2$, $\hat{m}_{LB}$, $\hat{m}_{MG}$ and $d_{cor}$ were used to estimate the ID of four real datasets from the UCI machine learning repository: (1) Housing values in suburbs of Boston, (2) Statlog (vehicle silhouettes), (3) Statlog (image segmentation) and (4) Combined cycle power plant. The output variables and the duplicate events were removed. The number of points $N$ and the number of variables $E$ of the resulting datasets are indicated in Table \ref{Real_dat_tab}. Notice that the true ID of real data is unknown and the goal of this subsection is to show that the results provided by $M_2$ are coherent with the theory and with the observations made for the synthetic data.\\ 

\begin{figure}[H]
\begin{center}
\includegraphics[width=13.5cm]{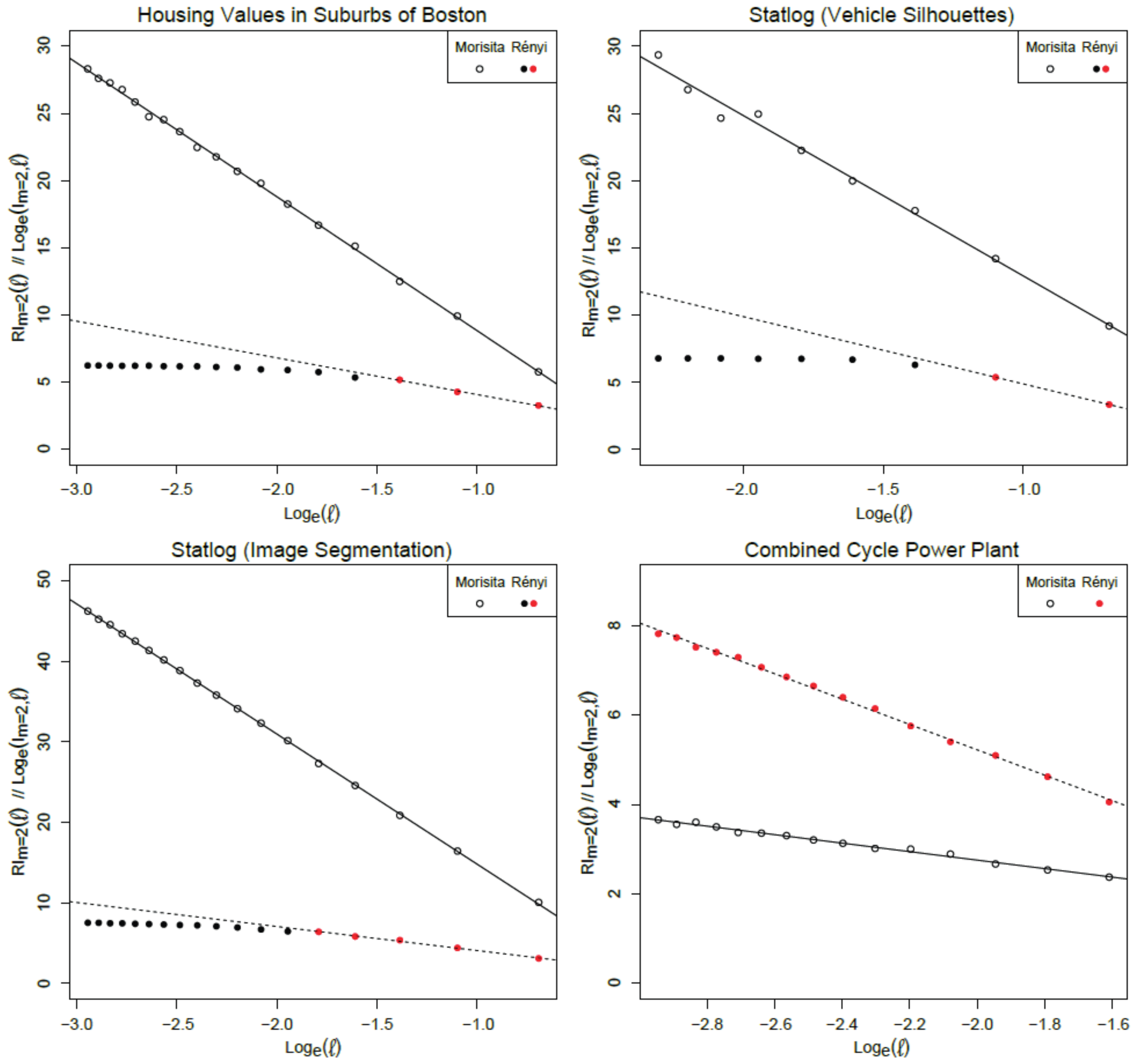}
\end{center}
\caption{Application of $I_{2,\ell}$ and $RI_{2}(\ell)$ to four real datasets}
\label{RealData}
\end{figure}

Figure \ref{RealData} shows the linear regressions required for the computation of $M_2$ and $D_2$. Here as well, $RI_2(\ell)$ appears to converge to a constant value as $\ell$ decreases. The departure from the power law denoted by the dashed line is more pronounced when $N$ is relatively low and the assumed ID rather high (according to the ID estimates of Table \ref{Real_dat_tab}). This can be observed with the vehicle silhouette data. In contrast, the power plant dataset is characterized by a great number of points $N$ and a relatively low ID (according to the final ID estimates of Table \ref{Real_dat_tab}). As a result, the entire range of the parameter $\ell$ can be used to perform linear regression. But, regarding $M_2$, in each of the case studies, the solid line is a good approximation of the behaviour of $\log{(I_{2,\ell})}$ and all the values of $\ell$ can be systematically retained for ID estimation. This primacy of $M_2$ over $D_2$ is in accordance with the results found for the synthetic data, for which the Morisita estimator always provided equivalent or better ID estimates than $D_q$. The variability of $\log{(I_{2,\ell})}$ for small values of $\ell$, observed in the case of the vehicle silhouettes, was also shown to have only a limited effect on the accuracy of $M_2$ (see Figures \ref{rnd_Cloud} and \ref{Cantor_set}). Based on this examination of Figure \ref{RealData}, $M_2$ appears to adequately capture the scaling properties of the four real world datasets.\\

\begin{table}
\centering
\scriptsize 
\begin{tabular}{|c||c|c|c|c|c|c|}
\hline 
Real Data & Housing Values & Vehicle Silhouettes & Image Segmentation & Power Plant\\ 
\hline
\hline 
$N$           & $506$  & $846$  & $2086$ & $9527$ \\ 
$E$           & $13$   & $18$   & $19$   & $4$   \\
\hline 
$M_2$       & \specialcell{$3.02$\\$\ell^{-1} \in\lbrace 2,3\dotsm,19\rbrace$} & \specialcell{$6.04$\\$\ell^{-1} \in\lbrace 2,3\dotsm,10\rbrace$} & \specialcell{$2.84$\\$\ell^{-1} \in\lbrace 2,3\dotsm,19\rbrace$} & \specialcell{$3.05$\\$\ell^{-1} \in\lbrace 5,6\dotsm,19\rbrace$}\\ 
\hline
$D_2$         & \specialcell{$2.73$\\$\ell^{-1} \in\lbrace 2,3,4\rbrace$} & \specialcell{$5.02$\\$\ell^{-1} \in\lbrace 2,3\rbrace$} & \specialcell{$2.99$\\$\ell^{-1} \in\lbrace 2,3,4,5,6\rbrace$} & \specialcell{$2.84$\\$\ell^{-1} \in\lbrace 5,6\dotsm,19\rbrace$}\\
\hline
$\hat{m}_{LB}$& \specialcell{$3.64$\\$k \in\lbrace 4,5,\dotsm,20\rbrace$} & \specialcell{$5.63$\\$k \in\lbrace 3,4,\dotsm,11\rbrace$} & \specialcell{$3.54$\\$k \in\lbrace 10,11,\dotsm,20\rbrace$} & \specialcell{$3.11$\\$k \in\lbrace 4,5,\dotsm,20\rbrace$}\\
\hline
$\hat{m}_{MG}$& \specialcell{$2.98$\\$k \in\lbrace 4,5,\dotsm,20\rbrace$} & \specialcell{$5.57$\\$k \in\lbrace 3,4,\dotsm,11\rbrace$} & \specialcell{$2.98$\\$k \in\lbrace 4,5,\dotsm,20\rbrace$} & \specialcell{$2.78$\\$k \in\lbrace 4,5,\dotsm,20\rbrace$}\\
\hline
$d_{cor}$     & \specialcell{$2.87$\\$0.05\leq\delta\leq0.5$} & \specialcell{$4.98$\\$0.1\leq\delta\leq0.4$} & \specialcell{$2.69$\\$0.04\leq\delta\leq 0.15$} &\specialcell{ $2.97$\\$0.02\leq\delta\leq 0.16$}\\ 
\hline
\end{tabular}
\caption{Application of the ID estimators to four real datasets from the UCI machine learning repository}
\label{Real_dat_tab}
\end{table}

The final ID estimates of the real world datasets are given in Table \ref{Real_dat_tab}. Although the results provided by $D_2$ were computed by using only the red points, they are less than those of $M_2$, except for the image segmentation data. It is also worth mentioning that $D_2$ yields $3.09$ for the power plant dataset if it is calculated with only the three points on the right side of the plot, which is close to the estimate of $M_2$. These observations are coherent with the conclusions drawn from the synthetic data: for the lowest considered values of $\ell$, the behaviour of $RI_2(\ell)$ can be evaluated by linear regression and the corresponding ID estimates tend to slightly underestimate both the ground truth and the values yielded by $M_2$ over the entire range of examined scales. Besides, the difference between the two estimators fades away for rather large $N$ and low theoretical ID (see Figures \ref{rnd_Cloud} and \ref{Cantor_set}). The power plant and the vehicle silhouette datasets are real world illustrations of this last point. The number $N$ of the former is large enough to fully characterize a 3-dimensional space (the assumed ID is about 3 and $N = 9527$), while the space where the latter resides is sparsely filled with points (the assumed ID is greater than 5 and $N = 846$). As a result, the range of $\ell$ and the final ID estimate of the two estimators differ more for the vehicle silhouette dataset than for the power plant one. In each plot of Figure \ref{RealData}, the difference is highlighted by the gap between the dashed line and the actual values of $RI_2(\ell)$ at the lowest scales. It is large in the case of the vehicle silhouettes, while it is non-existent for the power plant data. The two remaining datasets (i.e. the housing values and the image segmentation) are in an intermediate situation, as it is confirmed by the values given in Table \ref{Real_dat_tab}. Thus, the comparison between $M_2$ and $D_2$ using real world data shows an excellent consistency with what was expected from the theory and the synthetic data analysis.\\ 

Regarding the distance-based methods, the estimates of $df_{cor}$ are, as expected, lower than those provided by $M_2$. In contrast, the results of $\hat{m}_{LB}$ and $\hat{m}_{MG}$ tend to be less similar than for the synthetic data and they do not systematically underestimate the values of $M_2$. These slight differences, related to the complex specificities of the analysed datasets, do not question the good performance of the suggested estimator. Indeed, in each of the case studies, the values of $M_2$ are close to, at least, one of the two MLE-based results. And, finally, by comparison with the range of values yielded by all the other estimators, it can be definitely maintained that $M_2$ provides sensible ID estimates.

\section{Conclusion}\label{clus}
The Morisita estimator, $M_m$, is a new tool for estimating the Intrinsic Dimension (ID) of data. It is related to Rényi's generalized dimensions, $D_q$, for $m=q\geq 2$. $M_m$ tended to provide better results than $D_q$ on the synthetic data used in this study. This turned out to be particularly true for order 2 (i.e. $m=q=2$) when the data points were sparsely distributed. The application to four real datasets from the UCI machine learning repository confirmed the good properties of the suggested estimator. From the perspective of pattern recognition, $M_2$ might be of great interest, since it could be a good replacement for $D_2$ in algorithms, such as the fractal dimension algorithm \cite{Trai00,Trai10}. It might also open a new door to fractal supervised feature selection \cite{Mo12} of large datasets (our current work in progress), since its accuracy is coupled with a high computational efficiency.\\

$M_2$ was also compared with three distance-based estimators, namely $d_{cor}$, $\hat{m}_{LB}$ and $\hat{m}_{MG}$. It yielded good results when applied to the synthetic data and, by comparison, the real world applications revealed that it was able to provide reasonable ID estimates in challenging case studies. Therefore, in addition to the above-mentioned application in feature selection, the Morisita estimator of ID can be considered as a new tool for conducting advanced data mining tasks (e.g. dimensionality reduction \cite{LeeVer07}, monitoring network analysis \cite{lov86}) in many varied fields where the other estimators are commonly used (e.g. pattern recognition \cite{Cama03}, physics \cite{Grass831,Hent83}, cosmology \cite{Bor93}, climatology \cite{Lov13} and ecology \cite{Seur09}).\\

Finally, it is also worth mentioning that the multipoint Morisita index is a ratio of probabilities deep-rooted in the field of spatial clustering analysis. Consequently, $M_m$ can be viewed from a dual perspective, a fractal one and a statistical one, which helps to interpret the results.

\section{Acknowledgements}
The authors are grateful to the anonymous reviewers for their helpful and constructive comments that contributed to improving the paper. They also would like to thank Michael Leuenberger and Zhivko Taushanov for many fruitful discussions about machine learning and statistics.

\bibliographystyle{elsarticle-num}
\bibliography{References}
 
\end{document}